\documentclass[conference]{IEEEtran}
\IEEEoverridecommandlockouts
\usepackage{cite}
\usepackage{amsmath,amssymb,amsfonts}
\usepackage{graphicx}
\usepackage{textcomp}
\usepackage{subcaption}
\usepackage{xcolor}
\usepackage{multirow}
\usepackage{algorithm}
\usepackage{algpseudocode}
\def\BibTeX{{\rm B\kern-.05em{\sc i\kern-.025em b}\kern-.08em
    T\kern-.1667em\lower.7ex\hbox{E}\kern-.125emX}}
\begin{document}

\title{\textbf{tuGEMM}: Area-Power-Efficient Temporal Unary GEMM Architecture for Low-Precision Edge AI}
%
\author{\IEEEauthorblockN{%
    Harideep Nair,
    Prabhu Vellaisamy, 
    Albert Chen, Joseph Finn, Anna Li, Manav Trivedi, John Paul Shen
  }%
  \IEEEauthorblockA{Electrical and Computer Engineering Department, Carnegie Mellon University}%
}

\maketitle

\begin{abstract}

General matrix multiplication (GEMM) is a ubiquitous computing kernel/algorithm for data processing in diverse applications, including artificial intelligence (AI) and deep learning (DL). 
Recent shift towards edge computing has inspired GEMM architectures based on unary computing, which are predominantly stochastic and rate-coded systems.
%
This paper proposes a novel GEMM architecture based on temporal-coding, called \textit{tuGEMM}, that performs \textit{exact} computation. We introduce
two variants of \textit{tuGEMM}, \textit{serial} and \textit{parallel}, with distinct area/power-latency trade-offs. Post-synthesis Power-Performance-Area (PPA) in 45 nm CMOS are reported for 2-bit, 4-bit, and 8-bit computations. The designs illustrate significant advantages in area-power efficiency over state-of-the-art stochastic unary systems especially at low precisions, e.g. incurring just 0.03 mm\textsuperscript{2} and 9 mW for 4 bits, and 0.01 mm\textsuperscript{2} and 4 mW for 2 bits. This makes tuGEMM ideal for power constrained mobile and edge devices performing always-on real-time sensory processing.
%

\end{abstract}


\section{Introduction and Background}
General matrix multiplication (GEMM) 
performs multiply-and-accumulate operations on matrices and forms the fundamental building block for deep neural networks (DNNs). While the application performance of DNNs has increased steadily over the years, its computational demands have increased exponentially \cite{thompson2020computational}. Traditionally, GEMM was implemented as software libraries for CPUs and GPUs \cite{cuBLAS, CLBlast}. However, more recently, dedicated GEMM hardware units have been implemented within GPUs and DNN accelerators to improve compute efficiency. The increasing demand for hardware acceleration resulted in companies like Nvidia introducing the tensor cores \cite{yan2020demystifying} capable of performing 4x4 matrix multiplication,
and Google introducing Tensor Processing Units (TPUs) \cite{jouppi2017datacenter} with Matrix Multiply Units (MXUs). Further, with the latest push towards edge computing and on-device AI \cite{shi2016edge, li2019edge}, focus has shifted towards developing low-footprint GEMM hardware.
Towards that goal, Nvidia and Google introduced Jetson Xavier NX \cite{ditty2018nvidia}, and edge TPU \cite{cass2019taking} respectively, both of which deploy reduced compute on device, albeit at the expense of inference accuracy. 
In the current landscape, various such lightweight systems have been proposed \cite{pilipovic2021approximate, liu2020systolic}, including deep learning accelerators (DLAs) in modern smartphones \cite{ignatov2018ai}. These systems predominantly operate on binary values and trade off inference accuracy to meet the Power-Performance-Area (PPA) constraints for edge devices. 

On the other hand, \textit{unary} compute-based implementations offer a promising alternative solution to the increasing parallel computation complexity inherent to binary implementations, delivering low area and low power designs. This emerging paradigm replaces the multiple parallel bits with a single serial bit-stream.
Unary computing manifests in two major forms, \textit{rate} and \textit{temporal} coding, with rate-based methods being more prevalent. Recently proposed \textit{uGEMM} \cite{wu2020ugemm} is a unified rate-and-temporal encoded GEMM architecture that incorporates unary arithmetic units to perform stochastic GEMM operations. It provides significant PPA improvements compared to previous unary designs, while maintaining high accuracy, making it a promising candidate for edge devices. However, being a stochastic approach, it doesn't perform exact compute and falls under the widely researched domain of approximate computing. In contrast, our work focuses on exact, not approximate, GEMM compute based purely on temporal encoding.

Recent works have emphasized the current trend of AI/DL to move towards lower precision.
Authors in \cite{miyashita2016convolutional} performed training with 5-bit weights and 4-bit activations, and in \cite{mellempudi2017mixed} with 4-bit weights and 8-bit activations, both with minimal accuracy degradation.
More recently, IBM researchers achieved 8-bit precision for training and 4-bit precision for inference across many deep learning datasets \cite{wang20188}, followed by the work in \cite{sun2020ultra} that shows both training and inference can be performed with 4-bits with negligible impact on accuracy. Additionally, Akida NSoCs \cite{akida} employ 1, 2, 4-bit computations for their weights and activations targeting edge inference. This growing affinity towards low precision influences this work to explore low bit-width implementations for edge AI.

We propose a novel GEMM architecture, \textit{tuGEMM}, based on exact temporal compute, targeting area-power efficiency for low precision edge AI.
%
Our key contributions are as follows:
\begin{itemize}
    \item We propose a novel temporal-coding-based GEMM architecture that performs exact computations in contrast to existing stochastic and rate-based approaches.
    \item Two architecture variations, \textit{serial} and \textit{parallel}, that offer different area-latency tradeoffs are introduced.
    \item Gate-level implementations for both designs are presented, and post-synthesis PPA numbers for 2-bit, 4-bit and 8-bit implementations are reported.
    \item Latency evaluation for tuGEMM compute is performed using a representative DNN workload, ResNet18.
    \item We illustrate superior area-power efficiency over state-of-the-art unary approach, especially for low precision.
\end{itemize}

Section \ref{Sec:Arch} presents the encoding mechanism and the architecture of the two design variants. Section \ref{Sec:Eval} evaluates tuGEMM against uGEMM based on latency and post-synthesis area and power. Conclusions and future directions are  in Section \ref{Sec:Conc}.

\section{tuGEMM Architecture}
\label{Sec:Arch}

General matrix multiplication (GEMM) is central to the compute-intensive operations in DNNs. Its general format is:
\begin{align}
    \textbf{Y} = \alpha \textbf{AB} +\beta \textbf{C}
\end{align}
where $\textbf{A}$, $\textbf{B}$ and $\textbf{C}$ are generic $M$x$N$, $N$x$P$ and $M$x$P$ input matrices, $\textbf{Y}$ is the $M$x$P$ output matrix, and $\alpha$ and $\beta$ are scaling factors. This work focuses on \textit{non-scaled} GEMM operation, i.e., $\alpha$ = $\beta$ = 1.
%
%
%
This section describes the input encoding and the micro-architecture of the proposed \textit{tuGEMM} hardware.
%
\subsection{Input Encoding}
The key idea of unary hardware implementation is encoding values as serial bitstreams on a single bitline. Such an input encoding allows the hardware to be repurposed with significantly less area and power.
Unary encoding can be accomplished in two ways: rate and temporal coding. Rate-based systems encode values in the frequency of ones randomly distributed across the bitstream, whereas temporal coding encodes values in the time duration for which a signal is asserted. As a result, a temporally encoded bitstream consists of consecutive ones followed by consecutive zeros, resulting in only two transitions. This naturally leads to improved dynamic power consumption, compared to rate coding with multiple signal transitions due to the distributed occurrence of ones.
%
%

Rate coding typically implements stochastically-generated bitstreams using expensive random number generators (RNGs)
and suffers from the correlation problem, requiring additional hardware to mitigate it \cite{liu2017energy, lee2018correlation, alaghi2013exploiting}. In contrast, temporal encoding uses a single contiguous $n$-cycle wide logic pulse to represent a value $n$, analogous to the spike encoding employed in neuromorphic computing \cite{nair2021online}, and can enable exact deterministic compute in an efficient manner as it does not require RNGs. Our proposed approach distinguishes from previous works by utilizing temporal-unary-encoded exact compute. 
%

%
%

\subsection{Serial Architecture}
%


\begin{figure}[t]
\centering
\includegraphics[width=0.99\columnwidth]{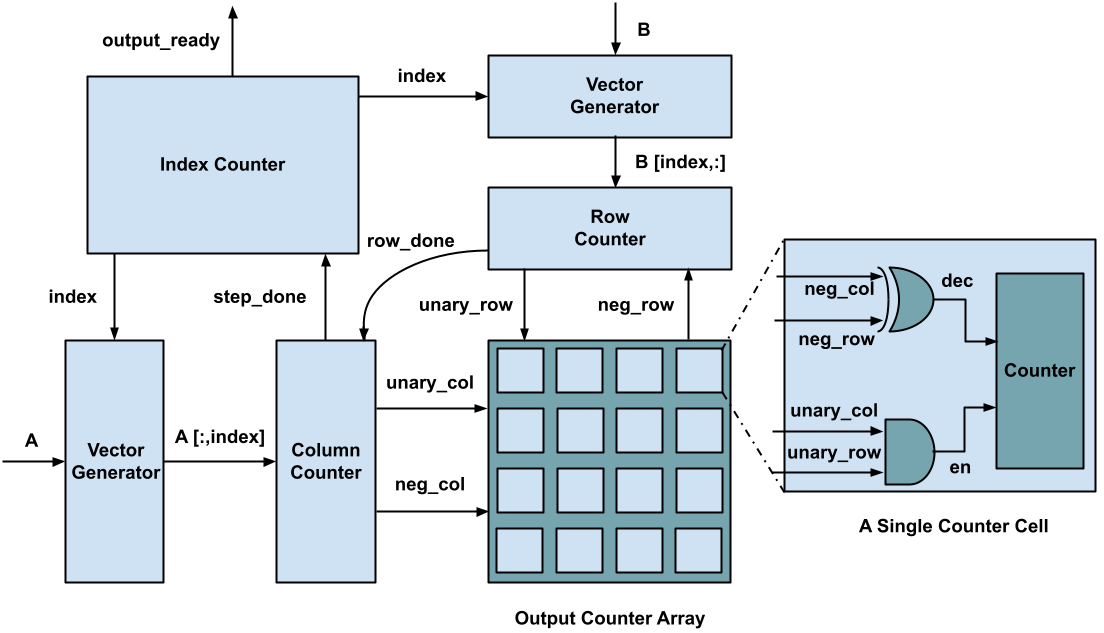} 
\caption{Serial tuGEMM architecture for 4x4 GEMM compute}
\label{arch1}
\end{figure}


The serial architecture (Fig. \ref{arch1}) consists of an $M$x$P$ array of \textit{output counter cells}
surrounded by peripheral logic that performs unary encoding and co-ordinates the dataflow into the array. The counter array receives unary-encoded input matrices \textbf{A}, \textbf{B} from the left and top respectively, and implements multiplication in unary fashion. The multiplication compute occurs in $N$ \textit{steps}, where $N$ is the common matrix dimension, which is equal to the number of columns in \textbf{A} and number of rows in \textbf{B}. Each \textit{step} computes the outer product of \textit{i}\textsuperscript{th} column from \textbf{A} and \textit{i}\textsuperscript{th} row from \textbf{B}. During each \textit{column-row} outer product, the $M$x$P$ output counters update their counts with the $M$x$P$ output values, taking as many cycles as the magnitude of the maximum output value (due to unary multiplication which will be described shortly). Thus, these outer products are accumulated over the $N$ \textit{steps}, at the end of which the final counter values reflect the output matrix \textbf{Y}. To eliminate a separate adder, the $M$x$P$ counters are initialized with the binary-encoded input matrix \textbf{C}
(following sections focus only on \textbf{A} \textbf{x} \textbf{B} multiplication).
%
The serial architecture performs the $N$ \textit{steps} serially, and is named so. 
It has four components:
\subsubsection{index counter} Each column of \textbf{A} is indexed simultaneously with the corresponding row of \textbf{B}. This indexing is generated by the \textit{index counter} that counts up from $0$ to $N-1$, incrementing each time by one after every \textit{step}, indicated by \textit{step\_done} signal. Once its count reaches $N$, the index counter asserts an \textit{output\_ready} signal, implying GEMM has finished.
%
\subsubsection{vector generator} Two vector generators receive index $i$ from the \textit{index counter} and use it to index into the input matrices \textbf{A} and \textbf{B} to generate 
the $i^{th}$ column of \textbf{A} ($M$-dimensional vector), and $i^{th}$ row of \textbf{B} ($P$-dimensional vector).
%
\subsubsection{column/row counters} 
$M$ \textit{column counters} and $P$ \textit{row counters} convert the binary values from the vector generator to unary signals and co-ordinates the unary multiplication.
In every \textit{step}, the column and row values from the \textit{vector generator} are loaded into the counters, which then begin counting towards zero (decrement if the initialized count is positive, increment if negative). The counters operate in a nested fashion such that 
the row counters are updated by 1 every cycle whereas the column counters only update their values (by 1) once all row counters reach zero. This cycle repeats until all the column counters reach zero thus completing one \textit{step}, eventually triggering the \textit{index counter} for the next \textit{step} via \textit{step\_done} signal. During every \textit{step}, the column and row counters assert $M$ \textit{unary\_col} and $P$ \textit{unary\_row} signals respectively, whenever their corresponding counts are non-zero. These signals represent the converted unary signals derived from the vector generator values, and enable column-row outer product in the \textit{output array} as will be described next. Each counter also asserts a \textit{neg\_col}/\textit{row} signal, if the corresponding initialized count is negative, to determine the direction of update in the output counter cells.
\subsubsection{output counter array}
It consists of $M$x$P$ counter cells (initialized with matrix \textbf{C}), where each counter accumulates the unary \textit{column-row} outer product within a \textit{step}, and is enabled when \textit{unary\_col}/\textit{row} are both asserted.
If enabled, it increments every cycle if the inputs have the same sign, else it decrements.
When the \textit{index counter} asserts \textit{output\_ready}, the output counter array holds \textbf{AB} + \textbf{C}. Note that the final result is binary, which enables direct cascading of multiple tuGEMM units as input values to the vector generators are binary.
\begin{figure}[t]
\centering
\includegraphics[width=0.91\columnwidth]{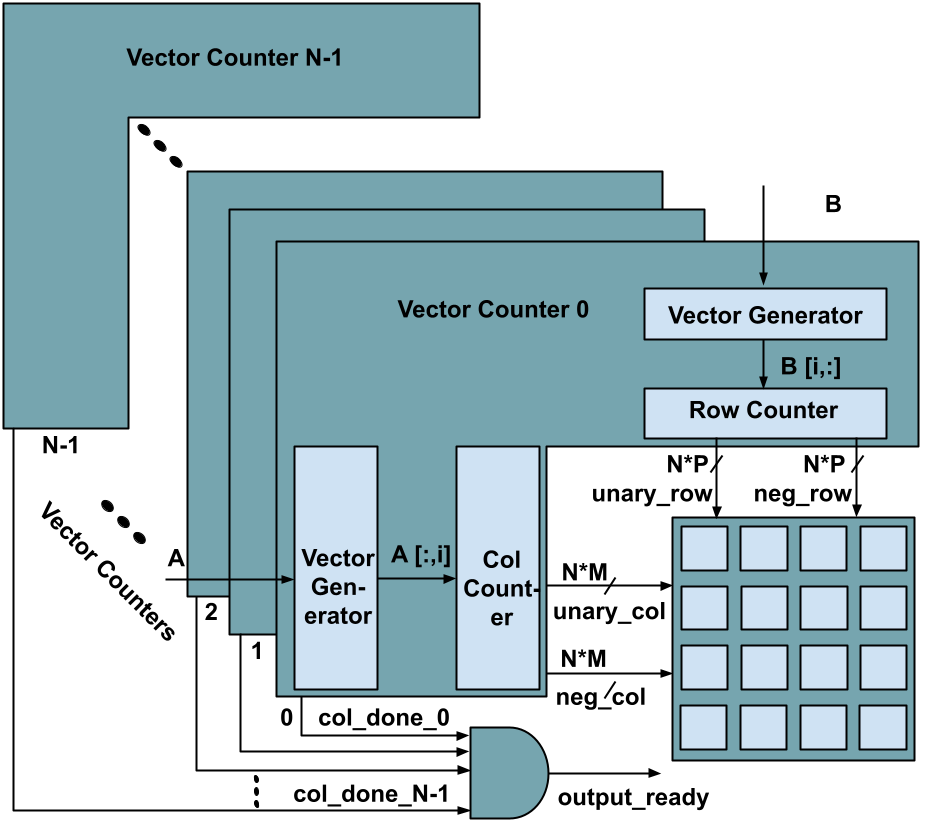} 
\caption{Parallel tuGEMM architecture for 4x4 GEMM compute}
\label{arch2bd}
\end{figure}
\subsection{Parallel Architecture}
A key observation is that the $N$ computation \textit{steps} are independent of each other. Hence, unlike serial, the parallel architecture (Fig. \ref{arch2bd}) computes all the $N$ \textit{steps} in parallel giving it its name, and is designed for reduced latency at the cost of increased area and power. To achieve this, it integrates the two \textit{vector generators}, and the column and row counters into a single \textit{vector counter} that is replicated $N$ times. It also houses an $M$x$P$ array of \textit{output adder cells} instead of the output counter array as in serial architecture, where each cell is now capable of adding the counts from all $N$ steps in parallel. Note that there is no need for an index counter used earlier to serialize the $N$ steps. The two main components here are:
%
%
\begin{figure}[t]
\centering
\includegraphics[width=0.95\columnwidth]{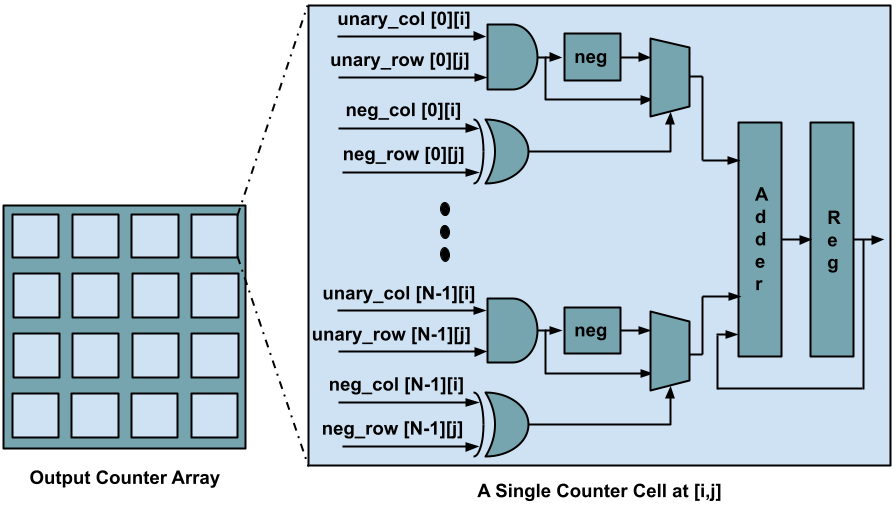} 
\caption{A single output adder cell in parallel tuGEMM}
\label{arch2c}
\end{figure}
\subsubsection{vector counters}
The $i^{th}$ \textit{vector counter} generates unary signals for $i^{th}$ column from \textbf{A} and $i^{th}$ row from \textbf{B}. Once all the $M$ column counters within a vector counter reach zero, it asserts its \textit{col\_done} signal. The GEMM computation finishes when all the vector counters assert this signal, generating \textit{output\_ready} signal via an AND gate. The vector counters output $N$ sets of $M$-dimensional \textit{unary} and \textit{neg} signals from the left and $N$ sets of $P$-dimensional \textit{unary} and \textit{neg} signals from the top, to be used by the output adder array.
\subsubsection{output adder array}
This component (Fig. \ref{arch2c}) holds majority of the hardware modifications with respect to the serial design. Firstly, the counter in the output counter cell of serial architecture is replaced by an adder and a register for accumulation, as necessitated by the requirement for computing all \textit{steps} in parallel. Secondly, each \textit{output adder cell} is capable of processing $N$ different pairs of \textit{unary} and \textit{neg} signals in parallel. Each of the $N$ pairs generates `1', `-1' (\textit{neg} block is used to generate `-1' in two's complement) or `0' based on the \textit{unary\_col/row} and \textit{neg\_col/row} signals controlled using a simple multiplexer,
that are fed into a binary adder which accumulates into a register which holds final value for that cell in the GEMM output, after \textit{output\_ready} is asserted.

\section{Evaluation and Results}
\label{Sec:Eval}
%

\subsection{Post-Synthesis Area-Power Evaluation}
Serial and parallel tuGEMM designs are implemented in System Verilog, and synthesized with Nangate45 Open Cell Library using Synopsys Design Compiler. Post-synthesis area and power are compared against uGEMM \cite{wu2020ugemm} for 8-bit 16x16 matrices ($M$=$N$=$P$=16) at 400 MHz (uGEMM uses this configuration). We further extend the design space to larger 32x32 matrices and lower precision (4 bits and 2 bits).

\begin{figure}[t]
 \centering
 \includegraphics[width=0.99\columnwidth]{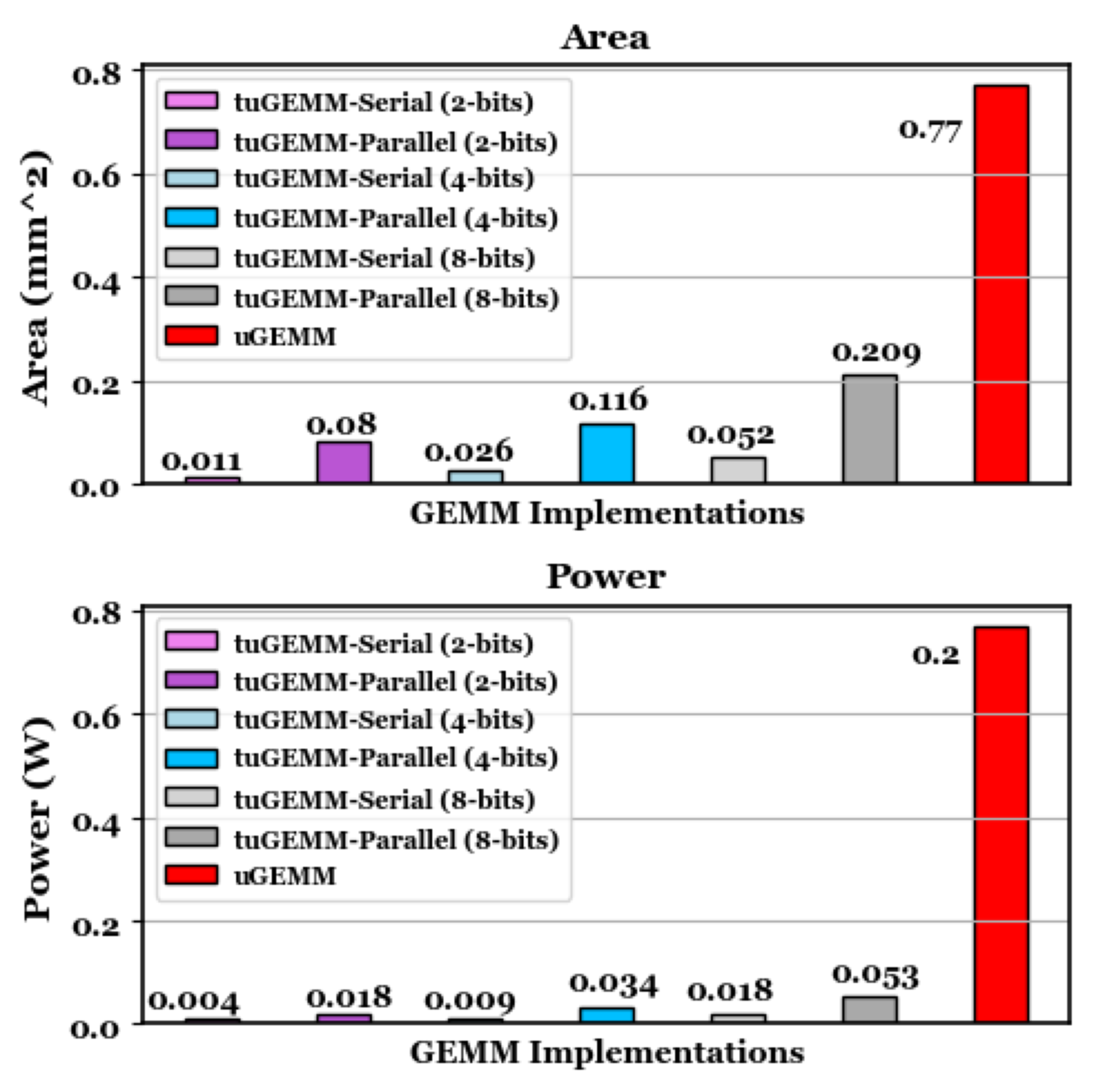} 
 \caption{PPA comparison for 16x16 GEMM implementations (serial/parallel tuGEMM vs. uGEMM) across 2, 4, 8 bitwidths.}
 \label{ppa_compare}
\end{figure}



All post-synthesis tuGEMM numbers are reported in Table \ref{tab:bit}, with 16x16 values illustrated in Fig. \ref{ppa_compare}, alongside uGEMM. It can be observed that
%
%
both serial and parallel tuGEMM are significantly more area-power efficient with respect to uGEMM. The parallel design consumes 3.7x and 3.8x less area and power, respectively, while the serial design reduces them by 14.8x and 11.1x, respectively. Also, serial design incurs 5.2x and 3.7x less area and power than parallel design.
Works in \cite{sun2020ultra, mellempudi2017mixed, miyashita2016convolutional, wang20188} suggest very low bit-widths are sufficient for DNNs to perform training and inference without significantly affecting the accuracy. tuGEMM shows significant PPA benefits in transitioning to very low bit-widths. On average, for every 2x reduction in bit-width, the area, power and delay are reduced by 2.1x, 2x, and 1.2x for serial, and 1.6x, 1.7x, and 1.1x for parallel designs, respectively.

\begin{table}[t]
\centering
\caption{45nm post-synthesis tuGEMM area-power (16x16 and 32x32 for 2, 4, 8 bits). 16x16 uGEMM baseline included.}
\scalebox{1}{
 \begin{tabular}{|c|c||c|c||c|c|} 
 \hline
 GEMM & Bit- & Area & Power & Area & Power \\
 Hardware & Width & (mm\textsuperscript{2}) & (W) & (mm\textsuperscript{2}) & (W)\\
 \cline{3-6}
 &  &  \multicolumn{2}{c||} {16x16} & \multicolumn{2}{c|} {32x32} \\
 \hline
 tuGEMM (serial) & 2 & 0.011 & 0.004 & 0.044 & 0.016  \\
 \hline
 tuGEMM (parallel) & 2 & 0.080 & 0.018 & 0.347 & 0.083 \\
 \hline
 tuGEMM (serial) & 4 & 0.026 & 0.009 & 0.099 & 0.034 \\
 \hline
 tuGEMM (parallel) & 4 & 0.116 & 0.034 & 0.506 & 0.145 \\
 \hline
 tuGEMM (serial) & 8 & 0.052 & 0.018 & 0.198 & 0.068 \\ 
 \hline
 tuGEMM (parallel) & 8 & 0.209 & 0.053 & 0.794 & 0.202 \\
 \hline
 \hline
 uGEMM (baseline) & 8 & 0.770 & 0.200 & - & - \\
 \hline
 \end{tabular}
 }
  \label{tab:bit}
\end{table}

Scaling matrix sizes, area and power for 32x32 tuGEMM increase by 4x compared to 16x16, as expected. An interesting observation here is that 32x32 parallel tuGEMM incurs similar area and power as 16x16 uGEMM (both 8-bit); 32x32 serial tuGEMM is more than 3x area-power efficient than 16x16 uGEMM. This superiority in area and power for tuGEMM arises by trading off latency, which is discussed in detail next.

\subsection{Latency Evaluation}
\label{seclat}
In this section, latencies for a complete GEMM compute are assessed. Assume $w$ is the input bit-width
for this discussion.
\subsubsection{Worst-Case Latency}
With two’s complement numbers, the largest  representable value is $2^{w-1}$. As the column/row counters perform unary encoding, it can take up to $2^{w-1}$ cycles for the row counter to reach zero and $(2^{w-1})*(2^{w-1})=(2^{w-1})^2$ cycles for the column counter to reach zero and generate the maximum value in any \textit{step}.
Since serial tuGEMM operates through $N$ such steps serially, it can take up to a maximum of $N*(2^{w-1})^2$ cycles in the worst case.
As can be seen from Table \ref{tab:bit}, the increase in area and power for parallel design compared to serial design is less than $N$-fold, potentially resulting in an overall boost in energy efficiency. 

Worst-case latency scales exponentially with bitwidth,
hence tuGEMM is best suited for low precision. 
However, average-case latencies for real workloads can be much lower depending on the frequency of large values.

\begin{figure}[t]
\centering
\includegraphics[width=0.93\columnwidth]{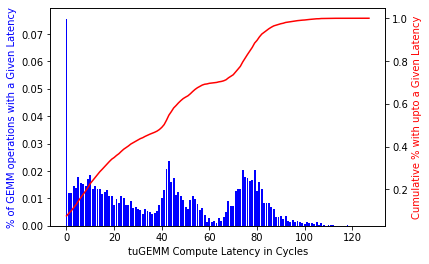} 
\caption{Percentage of GEMM operations that involve the corresponding X-axis values as the maximum magnitude during inference of INT8 quantized ResNet18. Right Y-axis plots the cumulative percentage of operations with maximum values less than or equal to the corresponding X-axis value.}
\label{percentf}
\end{figure}


\subsubsection{Average-Case Latency for Edge AI/DL}
Given the input matrices may not hold the largest absolute value, the actual latency can be significantly lower in most cases.
In order to profile maximum values, we use a representative edge DNN workload, INT8 quantized ResNet18.
During inference, in PyTorch, we keep track of the maximum values within each intermediate feature map and calculate the total number of times each value between 0 and 128 (maximum magnitude for 8-bit signed values) manifests as the maximum value within a feature map. 
This frequency of occurrence is plotted as percentage in Fig. \ref{percentf}. It illustrates that close to 8\% of the operations have 0 as the maximum value, about 50\% have maximum values less than 50 and 90\% have values less than 80. 
The average-case maximum value for ResNet18 can be calculated as area under the blue curve, which gives 41 (3x lower than 128). As a result, tuGEMM's average-case latency is significantly (10x) lower. 
This demonstrates the efficacy of tuGEMM in typical edge AI scenarios where much of the latency can be hidden due to sparsity of data values.
An accuracy evaluation on the same multi-layer perceptron from \cite{wu2020ugemm} yields 96.08\% for tuGEMM (exact) as opposed to 94.7\% for uGEMM (approximate). Exact compute becomes very important for lower precisions as any approximations can further exacerbate the quantization penalty on accuracy.

\section{Conclusion and Future Work}
\label{Sec:Conc}
This work introduces a novel temporal unary GEMM design, tuGEMM, capable of exact compute with very high area-power efficiency. In 45nm CMOS, 8-bit 16x16 serial tuGEMM consumes just 0.05 mm\textsuperscript{2} area and 18 mW power. The parallel design reduces serial latency by 16x while incurring only an increase of 5x/4x in area/power. Compared to state-of-the-art unary stochastic uGEMM, serial and parallel tuGEMM are about 15x/11x and 3.7x/3.8x more efficient in area/power respectively. This does incur a latency penalty which can be partially mitigated in tuGEMM by exploiting data sparsity and frequently occurring small values.
For 4-bit and 2-bit precisions, tuGEMM consumes minimal area and power with very reasonable latency, and thus can be excellent candidates for low-precision always-on edge-AI devices. 
%
%
Future research plans include
exploring different input encodings targeting latency optimization, and incorporating tuGEMM in DLAs.

\bibliographystyle{IEEEtranS}
\bibliography{refs}

\end{document}